\documentclass[
aps,
prd,
superscriptaddress,
amsfonts,
amssymb,
amsmath,
preprintnumbers,
floatfix,
nofootinbib,
unsortedaddress,
noshowpacs,
tightenlines,
floats,
letterpaper,
eqsecnum,
]{revtex4}

\begin{document}

\title[
Hamiltonian formulation of $ f(\text{Riemann}) $ theories of gravity
]{
Hamiltonian formulation of $ f(\text{Riemann}) $ theories of gravity
}

\author{Nathalie Deruelle}
\affiliation{
APC, UMR 7164 du CNRS, Universit\'e Paris 7,
75205 Paris, France
}

\author{Misao Sasaki}
\affiliation{
Yukawa Institute for Theoretical Physics, Kyoto University,
Kyoto 606--8502, Japan
}

\author{Yuuiti Sendouda}
\affiliation{
Yukawa Institute for Theoretical Physics, Kyoto University,
Kyoto 606--8502, Japan
}

\author{Daisuke Yamauchi}
\affiliation{
Yukawa Institute for Theoretical Physics, Kyoto University,
Kyoto 606--8502, Japan
}

\date{August 5, 2009}

\begin{abstract}
We present a canonical formulation of gravity theories whose Lagrangian
 is an arbitrary function of the Riemann tensor.
Our approach allows a unified treatment of various subcases and an easy
 identification of the degrees of freedom of the theory.
\end{abstract}

\preprint{YITP-09-47}

\maketitle

\section{
Introduction
}

Since H.\ Weyl introduced them in 1919
 \cite{Weyl1918ibWeyl:1919fiWeyl:1923}, theories of gravity whose
 action is nonlinear in the Riemann tensor (contrarily to Hilbert's)
 have been part of the ``landscape'' of General Relativity and its
 various extensions, up to present-day string theories.
In this paper we shall be interested in the action:
\begin{equation}
S_\mathrm g[g_{\mu\nu}]
= \frac{1}{2} \int_\mathcal M\!\mathrm d^Dx\,\sqrt{-g}\,
  f(\mathcal R_{\mu\nu\rho\sigma})\,.
\label{eq:action0}
\end{equation}
$ S_\mathrm g $ is a functional of the metric $ g_{\mu\nu}(x^\rho) $\,,
 $ D $ is the dimension of the spacetime $ \mathcal M $\,, and $ f $
 is an arbitrary function of the Riemann tensor
 $ \mathcal R_{\mu\nu\rho\sigma} $\,.
\footnote{
Our conventions are: $ \mathcal R_{\mu\nu\rho\sigma} =
 (1/2)\,(\partial_{\mu\rho} g_{\nu\sigma} -
 \partial_{\nu\sigma} g_{\mu\rho}) + \cdots $\,;
$ g $ is the determinant of $ g_{\mu\nu} $\,;
the signature is $ (-++\cdots) $\,;
spacetime indices $ (\mu,\nu,\cdots) $ run from $ 0 $ to $ D-1 $\,;
space indices $ (i,j,\cdots) $ will run from $ 1 $ to $ D-1 $\,.
}

The Euler--Lagrange equations of motion derived from metric variation
 of~\eqref{eq:action0} are
\begin{equation}
\mathcal R^{(\mu}{}_{\lambda\rho\sigma}\,
\frac{\partial f}{\partial\mathcal R_{\nu)\lambda\rho\sigma}}
- 2\,\nabla_\rho \nabla_\sigma
  \frac{\partial f}{\partial\mathcal R_{\rho(\mu\nu)\sigma}}
- \frac{1}{2}\,f\,g^{\mu\nu}
= T^{\mu\nu}\,,
\label{eq:eom0}
\end{equation}
where $ \nabla_\mu $ is the covariant derivative associated with
 $ g_{\mu\nu} $\,, $ T^{\mu\nu} $ is the energy-momentum tensor of matter,
 and $ f_{(\mu\nu)} \equiv (f_{\mu\nu} + f_{\nu\mu})/2 $\,.
These equations are generically fourth-order in the derivatives of the
 metric.
To convert them into a set of first-order differential equations, one
 must introduce as new variables some well-chosen functions of the
 metric and its derivatives.
The identification of these extra degrees of freedom (besides those of
 Einstein's gravity) is a prerequisite to introducing non-minimal
 coupling to matter.
It is also an important step to study for example the well-posedness
 of the initial value problem and the number of independent Cauchy
 data, the global charges associated with the solutions, the stability
 of the theory and the positivity of energy as well as the junction
 conditions.

At the linear approximation around $ D = 4 $ flat spacetime and in the
 case $ f =\alpha\,\mathcal R^2 +
 \beta\,\mathcal R_{\mu\nu}\,\mathcal R^{\mu\nu} $ the identification of
 the extra dynamical degrees of freedom was made long ago, see
 \cite{Stelle:1977ry}:
They are generally six, corresponding to a ``massive spin-$ 0 $''
 together with a ``massive spin-$ 2 $'' field (a ``ghost'' of negative
 energy), and reduce to five if $ f $ is the square of the Weyl tensor
 ($ \beta = -3\,\alpha $).
The action was then recast in first-order form and the constraints
 were analysed in \cite{Boulware:1983yj} and \cite{Buchbinder:1987vp}
 with the result, among others, that when $ f $ is the square of the
 Weyl tensor there is an additional, ``first-class,'' constraint which
 generates conformal transformations (see also
 \cite{Alonso:1994tr,Demaret:1998dm}).
The analysis of $ f = \alpha\,\mathcal R^2 +
 \beta\,\mathcal R_{\mu\nu}\,\mathcal R^{\mu\nu} +
 \gamma\,\mathcal R_{\mu\nu\rho\sigma}\,\mathcal R^{\mu\nu\rho\sigma} $ in $ D $
 dimensions was performed in \cite{Magnano:1990qu} with the result
 that the number of extra degrees of freedom is reduced if $ f $ is
 the square of the Weyl tensor ($ \alpha = 2 $\,, $ \beta =
 -4\,(D-1) $\,, $ \gamma = (D-1)\,(D-2) $) or the Gauss--Bonnet
 combination ($ \alpha = 1 $\,, $ \beta = -4 $\,, $ \gamma = 1 $).
Finally, a canonical formulation of $ f(\mathcal R) $ theories has
 been proposed in \cite{Ezawa:1998ax} (see also
 \cite{Deruelle:2009pu}) and the Lovelock case \cite{Lovelock:1971yv}
 was treated in \cite{Teitelboim:1987}.

Our aim here will be to unify these results and generalise them to the
 full action \eqref{eq:action0}.
To do so we shall follow the procedure of Arnowitt, Deser and Misner
 (ADM) \cite{Arnowitt:1962hi} and present a canonical formulation of
 theories of gravity yielding the field equations \eqref{eq:eom0}.
In contradistinction with previous approaches which consist in
 choosing as the extra variables either the extrinsic curvature of the
 ADM foliation \cite{Buchbinder:1987vp} or its time derivative
 \cite{Boulware:1983yj}, we shall ascribe a leading role to the
 components $ \mathcal R_i{}^0{}_j{}^0 \propto \Omega_{ij} $ of the
 Riemann tensor (in an ADM coordinate system adapted to the foliation).
This will allow us to write the canonical equations of motion in a
 compact form.
The extra degrees of freedom will be encoded in the $ D\,(D-1)/2 $
 components of some spatial tensor $ \Psi^{ij} $ and their number will
 depend on the number of components of $ \Omega_{ij} $ that can be
 extracted from the equation
\begin{equation}
2\,\Psi^{ij}
+ \frac{\partial f}{\partial\Omega_{ij}}
= 0\,.
\end{equation}

The paper is organised as follows.
In Section~\ref{sec:EL} we introduce a number of auxiliary fields in
 the action, perform its ADM decomposition and simplify it by using
 some of the constraints which simply determine some of the auxiliary
 fields algebraically.
This leads us to our second-order form Lagrangian, see \eqref{eq:L*}.
In Section~\ref{sec:H} we first obtain the Hamiltonian of the theory,
 see \eqref{eq:H} and \eqref{eq:C}, and, second, write Hamilton's
 equations of motion, see \eqref{eq:ceom1} and \eqref{eq:ceom2}.
Section~\ref{sec:red} presents the way to reduce the Hamiltonian using
 constraints and shows how a number of well-known subcases are
 recovered (General Relativity, $ f(\mathcal R) $ and
 ``$ \text{Weyl}^2 $'' theories).
Finally we make the link in an Appendix between our formalism and the
 commonly used Ostrogradsky one.

We shall present in this paper only the thread of the argument.
The details of the calculations will be presented elsewhere
 \cite{Sendouda:prep}.

\section{\label{sec:EL}
Choosing the action
}

\subsection{
Introducing auxiliary fields
}

In order to turn the equations of motion \eqref{eq:eom0} into a set of
 first-order differential equations we shall start our analysis, not
 with \eqref{eq:action0} but with the related action
\begin{equation}
S[g_{\mu\nu},\varrho_{\mu\nu\rho\sigma},\varphi^{\mu\nu\rho\sigma}]
= \frac{1}{2} \int_\mathcal M\!\mathrm d^Dx\,\sqrt{-g}\,
  [f(\varrho_{\mu\nu\rho\sigma})
   + \varphi^{\mu\nu\rho\sigma}\,
     (\mathcal R_{\mu\nu\rho\sigma} - \varrho_{\mu\nu\rho\sigma})]\,.
\label{eq:action2}
\end{equation}
The two auxiliary fields $ \varrho_{\mu\nu\rho\sigma} $ and
 $ \varphi^{\mu\nu\rho\sigma} $ have all the symmetries of
 $ \mathcal R_{\mu\nu\rho\sigma} $ and are chosen to be independent of each
 other and of $ g_{\mu\nu} $\,.
The reason for introducing them is that the second derivatives of the
 metric appear only linearly in \eqref{eq:action2} and will be
 eliminated by means of an integration by parts, see below.
If matter does not couple to $ \varphi^{\mu\nu\rho\sigma} $ and
 $ \varrho_{\mu\nu\rho\sigma} $\,, variation of \eqref{eq:action2} yields
 a set of field equations which is equivalent to \eqref{eq:eom0}.
Indeed:
\begin{equation}
\begin{aligned}
\delta S
& = \frac{1}{2} \int_\mathcal M\!\mathrm d^Dx\,\sqrt{-g}\,
    \left[
     -\mathcal E^{\mu\nu}\,\delta g_{\mu\nu}
     + (\mathcal R_{\mu\nu\rho\sigma} - \varrho_{\mu\nu\rho\sigma})\,
       \delta\varphi^{\mu\nu\rho\sigma}
     + \left(
        \frac{\partial f}{\partial\varrho_{\mu\nu\rho\sigma}}
        - \varphi^{\mu\nu\rho\sigma}
       \right)\,
       \delta\varrho_{\mu\nu\rho\sigma}
    \right] \\
& \quad
    + \int_\mathcal M\!\mathrm d^Dx\,\sqrt{-g}\,
      \nabla_\sigma
      [(\nabla_\rho \varphi^{\mu\rho\nu\sigma})\,\delta g_{\mu\nu}
        - \varphi^{\mu\rho\nu\sigma}\,(\nabla_\rho \delta g_{\mu\nu})]\,,
\label{eq:var}
\end{aligned}
\end{equation}
where
\begin{equation}
\mathcal E^{\mu\nu}
\equiv
  -\mathcal R^{(\mu}{}_{\alpha\beta\gamma}\,\varphi^{\nu)\alpha\beta\gamma}
  - 2\,\nabla_\alpha \nabla_\beta \varphi^{\alpha(\mu\nu)\beta}
  + 2\,\varrho^{(\mu}{}_{\alpha\beta\gamma}\,
    \frac{\partial f}{\partial\varrho_{\nu)\alpha\beta\gamma}}
  - \frac{1}{2}\,
    [\varphi^{\alpha\beta\gamma\delta}\,
     (\mathcal R_{\alpha\beta\gamma\delta} - \varrho_{\alpha\beta\gamma\delta})
     + f(\varrho_{\alpha\beta\gamma\delta})]\,
    g^{\mu\nu}\,.
\end{equation}
Ignoring the divergence term for the time being, the field equations
 are
\begin{equation}
\mathcal E^{\mu\nu}
= T^{\mu\nu}\,,
\label{eq:eom1}
\end{equation}
and, if the matter action does not depend on $ \varrho_{\mu\nu\rho\sigma} $
 and $ \varphi^{\mu\nu\rho\sigma} $\,,\footnote{
For an example of non-minimal coupling to matter, see
 e.g.\ \cite{Deruelle:2007pt,Deruelle:2008fs}.
}
\begin{equation}
\varrho_{\mu\nu\rho\sigma}
= \mathcal R_{\mu\nu\rho\sigma}\,,
\label{eq:eom2}
\end{equation}
together with
\begin{equation}
\varphi^{\mu\nu\rho\sigma}
= \frac{\partial f}{\partial\varrho_{\mu\nu\rho\sigma}}\,.
\label{eq:eom3}
\end{equation}
Substituting these extra equations into \eqref{eq:eom1}, we recover
 the original fourth-order equation of motion \eqref{eq:eom0}.

A remark is in order here.
As is well known \cite{Faddeev:1988qp}, one can plug the constraint
 \eqref{eq:eom3} back into the action \eqref{eq:action2} and consider
 it as a functional of $ g_{\mu\nu} $ and $ \varrho_{\mu\nu\rho\sigma} $ only.
Indeed variation of this new action still yields the same equations of
 motion, the only difference being that \eqref{eq:eom2} is replaced by
\begin{equation}
\frac{\partial^2f}{\partial\varrho_{\mu\nu\rho\sigma}\,
\partial\varrho_{\alpha\beta\gamma\delta}}\,
(\mathcal R_{\alpha\beta\gamma\delta} - \varrho_{\alpha\beta\gamma\delta})
= 0\,.
\label{eq:eom4}
\end{equation}
Whether this equation can be inverted to yield \eqref{eq:eom2} or not
 imposes an analysis of the various subcases \emph{before} any
 canonical treatment.
For this reason we shall refrain from using \eqref{eq:eom3}
 straightaway and shall stick to the action \eqref{eq:action2}.
This will allow us to treat the different subcases in a unified
 manner.

\subsection{
ADM decomposition
}
 
Suppose that $ \mathcal M $ can be foliated by a family of spacelike
 surfaces $ \Sigma_t $\,, defined by $ t = x^0 $\,.\footnote{
We here work in an ADM coordinate system.
See e.g.\ \cite{Misner:1974qy,Wald:1984rg,Gourgoulhon:2007ue} for a
 more geometrical approach.
}
Let $ h_{ij} \equiv g_{ij}|_{x^0=t} $ with $ i,j $ running from $ 1 $ to
 $ D-1 $ be the metric on $ \Sigma_t $\,, $ h $ its determinant,
 $ h^{ij} $ its inverse, and $ D_i $ be the associated covariant
 derivative.
Introduce the future-pointing unit normal vector $ n $ to the surface
 $ \Sigma_t $\,, that is, to the basis vector fields $ \partial_i $
 with components $ \delta^\mu_i $\,;
the components of $ n $ are $ n_i = 0 $\,,
 $ n_0 = -1/\sqrt{-g^{00}} $\,, $ n^0 = \sqrt{-g^{00}} $\,,
 $ n^i = -g^{0i}/\sqrt{-g^{00}} $\,.
Decompose then the timelike basis vector $ \partial_0 $ (with
 components $ \delta^\mu_0 $) on $ n $ and the spatial vectors
 $ \partial_i $\,:
$ \delta^\mu_0 = N\,n^\mu + \beta^i\,\delta^\mu_i $\,.
$ N = 1/\sqrt{-g^{00}} $ and $ \beta^i = -g^{0i}/g^{00}$ are the ``lapse''
 and ``shift,'' respectively;
together with the induced metric $ h_{ij} $\,, they constitute the
 ``ADM variables'' \cite{Arnowitt:1962hi}.
In terms of these variables we have
\begin{equation}
n_0
= -N\,,
\quad
n_i
= 0\,,
\quad
n^0
= \frac{1}{N}\,,
\quad
n^i
= -\frac{\beta^i}{N}\,.
\label{eq:n}
\end{equation}
The components of the spacetime metric read
\begin{equation}
\left\{
\begin{aligned}
&
g_{00}
= -N^2 + \beta_i\,\beta^i\,,
\quad
g_{0i}
= \beta_i\,,
\quad
g_{ij}
= h_{ij}\,, \\
&
g^{00}
= -\frac{1}{N^2}\,,
\quad
g^{0i}
= \frac{\beta^i}{N^2}\,,
\quad
g^{ij}
= h^{ij} - \frac{\beta^i\,\beta^j}{N^2}
\end{aligned}
\right.
\label{eq:ADM}
\end{equation}
and $ \sqrt{-g} = N \sqrt h $\,.\footnote{
Here and in the following indices of $ (D-1) $-dimensional spatial
 tensors are moved with the induced metric $ h_{ij} $\,.
}
Introduce finally the extrinsic curvature of $ \Sigma_t $\,:
\begin{equation}
K_{ij}
\equiv
  \nabla_i n_j
= \frac{1}{2\,N}\,
  (\dot h_{ij} - D_i \beta_j - D_j \beta_i)\,,
\label{eq:K}
\end{equation}
where a dot denotes a time derivative: $ \dot h_{ij} =
 \partial_t h_{ij} $\,.

A standard calculation using \eqref{eq:n}, \eqref{eq:ADM} and
 \eqref{eq:K} (see
 e.g.\ \cite{Misner:1974qy,Wald:1984rg,Gourgoulhon:2007ue} for a
 geometrical derivation) yields the Gauss, Codazzi and Ricci
 equations, that is, the components of the Riemann tensor in terms of
 the ADM variables:
\begin{equation}
\left\{
\begin{aligned}
\mathcal R_{ijkl}
& = K_{ik}\,K_{jl} - K_{il}\,K_{jk} + R_{ijkl}\,, \\
\mathcal R_{ijk\mathbf n}
& \equiv
    n^\mu\,\mathcal R_{ijk\mu}
  = D_i K_{jk} - D_j K_{ik}\,, \\
\mathcal R_{i\mathbf nj\mathbf n}
& \equiv
    n^\mu\,n^\nu\,\mathcal R_{i\mu j\nu}
  = -N^{-1}\,(\dot K_{ij} - \pounds_\beta K_{ij})
    + (K \cdot K)_{ij}
    + N^{-1}\,D_{ij} N\,,
\end{aligned}
\right.
\label{eq:GCRY}
\end{equation}
where $ R_{ijkl} $ is the Riemann tensor of $ h_{ij} $\,,
 $ (A \cdot B)_{ij} \equiv A_{ik}\,B_j{}^k $\,, $ D_{ij} \equiv
 D_i D_j $\,, and the short-hand notation $ \pounds_\beta $ denotes the
 Lie derivative with respect to the shift:
$ \pounds_\beta K_{ij} \equiv \beta^k D_k K_{ij} + K_{ik}\,D_j \beta^k +
 K_{jk}\,D_i \beta^k $\,.
We can thus perform the following decomposition:
\begin{equation}
\varphi^{\mu\nu\rho\sigma}\,
(\mathcal R_{\mu\nu\rho\sigma} - \varrho_{\mu\nu\rho\sigma})
= \phi^{ijkl}\,(\mathcal R_{ijkl} - \rho_{ijkl})
  - 4\,\phi^{ijk}\,(\mathcal R_{ijk\mathbf n} - \rho_{ijk})
  - 2\,\Psi^{ij}\,(\mathcal R_{i\mathbf nj\mathbf n} - \Omega_{ij})\,,
\end{equation}
where we have introduced the following spatial tensors evaluated on
 $ \Sigma_t $\,:
\begin{equation}
\left\{
\begin{aligned}
\rho_{ijkl}
& \equiv
    \varrho_{ijkl}\,,
\quad
\rho_{ijk}
  \equiv
    n^\mu\,\varrho_{ijk\mu}\,,
\quad
\Omega_{ij}
  \equiv
    n^\mu\,n^\nu\,\varrho_{i\mu j\nu}\,, \\
\phi^{ijkl}
& \equiv
    h^{im}\,h^{jn}\,h^{kp}\,h^{lq}\,\varphi_{mnpq}\,,
\quad
\phi^{ijk}
  \equiv
    h^{il}\,h^{jm}\,h^{kn}\,n^\mu\,\varphi_{lmn\mu}\,,
\quad
\Psi^{ij}
  \equiv
    -2\,h^{ik}\,h^{jl}\,n^\mu\,n^\nu\,\varphi_{k\mu l\nu}\,.
\end{aligned}
\right.
\label{eq:aux}
\end{equation}

In order to pass to canonical formulation, one must remove second
 derivatives from the action.
It is done by integrating the second derivative by parts to cast it
 into a total divergence.
A side exercise shows how the time derivative of the extrinsic
 curvature is transformed into a time derivative of $ \Psi^{ij} $\,:
\begin{equation}
N^{-1}\,\Psi^{ij}\,(\dot K_{ij} - \pounds_\beta K_{ij})
= \nabla_\mu (n^\mu\,K \cdot \Psi)
  - K\,(K \cdot \Psi)
  - N^{-1}\,K_{ij}\,(\dot\Psi^{ij} - \pounds_\beta \Psi^{ij})\,,
\label{eq:intp}
\end{equation}
where $ K \equiv h^{ij}\,K_{ij} $ and $ A \cdot B \equiv A_{ij}\,B^{ij} $\,.

Armed with these preliminaries we can decompose the action
 \eqref{eq:action2} as
\begin{equation}
S[h_{ij},N,\beta^i,\Psi^{ij},\Omega_{ij},
\phi^{ijkl},\phi^{ijk},\rho_{ijkl},\rho_{ijk}]
= \int_\mathcal M\!\mathrm d^Dx\,
  [\mathcal L + \partial_\mu (\sqrt{-g}\,n^\mu\,K \cdot \Psi)]\,,
\label{eq:S}
\end{equation}
with
\begin{equation}
\begin{aligned}
\mathcal L
& = \sqrt h\,N
    \biggl[
     \frac{1}{2}\,f(\varrho_{\mu\nu\rho\sigma})
     + \frac{1}{2}\,\phi^{ijkl}\,(\mathcal R_{ijkl} - \rho_{ijkl})
     - 2\,\phi^{ijk}\,(\mathcal R_{ijk\mathbf n} - \rho_{ijk}) \\
& \quad\quad
     - \Psi^{ij}\,(K\,K_{ij} + (K \cdot K)_{ij} + N^{-1}\,D_{ij} N - \Omega_{ij})
     - N^{-1}\,K_{ij}\,(\dot\Psi^{ij} - \pounds_\beta \Psi^{ij})
    \biggr]\,,
\label{eq:L}
\end{aligned}
\end{equation}
where $ \mathcal R_{ijkl} $ and $ \mathcal R_{ijk\mathbf n} $ are given in
 \eqref{eq:GCRY} and where it is understood that
 $ f(\varrho_{\mu\nu\rho\sigma}) $ is expressed in terms of $ \rho_{ijkl} $\,,
 $ \rho_{ijk} $ and $ \Omega_{ij} $ (see below for examples of such a
 decomposition).

As one can see from \eqref{eq:L} and \eqref{eq:GCRY} the spatial
 tensors $ \phi^{ijkl} $ and $ \phi^{ijk} $ are not dynamical and their
 equations of motion ($ \partial\mathcal L/\partial\phi^{ijkl} = 0 $\,,
 $ \partial\mathcal L/\partial\phi^{ijk} = 0 $) are simple constraints:
\begin{equation}
\rho_{ijkl}
= K_{ik}\,K_{jl} - K_{il}\,K_{jk} + R_{ijkl}\,,
\quad
\rho_{ijk}
= D_i K_{jk} - D_j K_{ik}\,,
\label{eq:cons1}
\end{equation}
which can harmlessly be incorporated into $ \mathcal L $\,, see
 \cite{Faddeev:1988qp}.
As a consequence the Lagrangian density of our theory reduces to
\begin{equation}
\mathcal L^*
= \sqrt h\,N\,
  \left[
   \frac{1}{2}\,f(\varrho_{\mu\nu\rho\sigma})
   - \Psi^{ij}\,(K\,K_{ij} + (K \cdot K)_{ij} + N^{-1}\,D_{ij} N - \Omega_{ij})
   - N^{-1}\,K_{ij}\,(\dot\Psi^{ij} - \pounds_\beta \Psi^{ij})
  \right]\,,
\label{eq:L*}
\end{equation}
where the spatial tensors $ \rho_{ijkl} $ and $ \rho_{ijk} $ in
 $ f(\varrho_{\mu\nu\rho\sigma}) $ are given by \eqref{eq:cons1}.

The divergence in \eqref{eq:S} is canceled by adding to the action
 the generalisation of the York--Gibbons--Hawking term
 \cite{York:1972sj,Gibbons:1976ue} (see also \cite{Hawking:1984ph})\,:
\begin{equation}
\bar S
= -\oint_{\partial\mathcal M}\!\mathrm d\Sigma_\mu\,n^\mu\,\Psi \cdot K\,,
\label{eq:surface}
\end{equation}
where $ \mathrm d\Sigma_\mu $ is the normal to the boundary
 $ \partial\mathcal M $ being proportional to the volume element of
 $ \partial\mathcal M $\,.
Here $ \mathrm d\Sigma_\mu $ is outward-pointing if spacelike and
 inward-pointing if timelike.
The field equations \eqref{eq:eom1}, \eqref{eq:eom2} and
 \eqref{eq:eom3} then derive from a Dirichlet variational principle
 where the induced metric $ h_{ij} $ and $ \Psi^{ij} $ have to be held
 fixed on $ \partial\mathcal M $\,.

Let us note that there can be constraints on $ \Psi^{ij} $ depending on
 the form of $ f $ (see Section~\ref{sec:eom} for general discussion
 about this constraint) and in some specific cases $ \Psi^{ij} $ is
 given as a function of $ K_{ij} $\,.
Then the surface term \eqref{eq:surface} cannot be used to remove the
 second derivatives from the action.
This is for example what happens in the Lovelock case
 \cite{Sendouda:prep}.
For simplicity, we do not consider such exceptional cases in this
 paper and the divergence in \eqref{eq:S} is hereafter discarded.
More details about the surface term will be presented elsewhere
 \cite{Sendouda:prep}.

\section{\label{sec:H}
Canonical formalism
}

\subsection{
First-order action and Hamiltonian
}

Momenta conjugate to the dynamical variables $ h_{ij} $ and $ \Psi^{ij} $
 are defined as, recalling the definition \eqref{eq:K} of $ K_{ij} $\,,
\begin{equation}
\left\{
\begin{aligned}
p^{ij}
  \equiv
    \frac{\delta\mathcal L^*}{\delta\dot h_{ij}}
& = -\frac{\sqrt h}{2}\,
    \biggl[
     h^{ij}\,\Psi \cdot K
     + K\,\Psi^{ij}
     + 2\,(\Psi \cdot K)^{(ij)}
     + N^{-1}\,(\dot\Psi^{ij} - \pounds_\beta \Psi^{ij})
     - 2\,\frac{\partial f}{\partial\rho_{ikjl}}\,K_{kl} \\
& \quad\quad
     + N^{-1}\,D_k \left(N\,\frac{\partial f}{\partial\rho_{k(ij)}}\right)
    \biggr]\,, \\
\Pi_{ij}
  \equiv
    \frac{\delta\mathcal L^*}{\delta\dot\Psi^{ij}}
& = -\sqrt h\,K_{ij}\,,
\end{aligned}
\right.
\end{equation}
where it is understood that the derivatives
 $ \partial f/\partial\rho_{ijkl} $ and $ \partial f/\partial\rho_{ijk} $
 have all the symmetries of $ \rho_{ijkl} $ and $ \rho_{ijk} $\,,
 respectively, and we have used the constraints \eqref{eq:cons1} and
 discarded a total divergence in $ p^{ij} $\,.\footnote{
When integrated over space this divergence yields the following
 contribution to the momentum: 
$ (1/2)\,\oint_S\!\mathrm dS_k\,(\partial f/\partial\rho_{k(ij)}) $\,,
 where $ \mathrm dS_k $ is the outward-pointing normal to the
 $ (D-2) $-dimensional sphere $ S $ being proportional to the
 volume element of $ S $\,.
Such a term, as well as all spatial divergences, will be discarded in
 this paper but are important when studying, e.g., the energy of the
 system or junction conditions \cite{Sendouda:prep}.
}
Inversion yields the velocities in terms of the canonical variables
 (see \eqref{eq:ceom1} below for their explicit expression) and the
 first-order Lagrangian then is
\begin{equation}
\mathcal L^*
= p \cdot \dot h
  + \Pi \cdot \dot\Psi
  - \mathcal H^*
  - \partial_i (\sqrt h\,V^i)
\label{eq:L*2}
\end{equation}
with
\begin{equation}
V^i
= \Psi^{ij}\,\partial_j N
  - N\,D_j \Psi^{ij}
  + 2\,\frac{p^{ij}}{\sqrt h}\,\beta_j
  - 2\,\frac{\Pi_{jk}}{\sqrt h}\,\Psi^{ij}\,\beta^k\,,
\end{equation}
where $ \mathcal H^* $ is the Hamiltonian density:
\begin{equation}
\mathcal H^*[h_{ij},p^{ij},\Psi^{ij},\Pi_{ij},N,\beta^i,\Omega_{ij}]
= N\,\mathcal C + \beta^i\,\mathcal C_i
\label{eq:H}
\end{equation}
with
\begin{equation}
\left\{
\begin{aligned}
\mathcal C
& \equiv
    \frac{1}{\sqrt h}\,
    ((\Psi \cdot \Pi)\,\Pi
     + \Psi \cdot \Pi \cdot \Pi
     - 2\,p \cdot \Pi)
    + \sqrt h\,
      \left[
       D_{ij} \Psi^{ij}
       - \left(
          \Psi \cdot \Omega + \frac{1}{2}\,f(\varrho_{\mu\nu\rho\sigma})
         \right)
      \right]\,, \\
\mathcal C_i
& \equiv
    -2\,\sqrt h\,D_j \left(\frac{p_i{}^j}{\sqrt h}\right)
    + \Pi_{jk}\,D_i \Psi^{jk}
    + 2\,\sqrt h\,D_k \left(\frac{\Pi_{ij}}{\sqrt h}\,\Psi^{jk}\right)\,,
\end{aligned}
\right.
\label{eq:C}
\end{equation}
where $ \Pi \equiv h^{ij}\,\Pi_{ij} $ and $ A \cdot B \cdot C =
 A^i{}_j\,B^j{}_k\,C^k{}_i $\,, and the spatial tensors $ \rho_{ijkl} $
 and $ \rho_{ijk} $ in $ f(\varrho_{\mu\nu\rho\sigma}) $ are now given by 
\begin{equation}
\rho_{ijkl}
= \frac{\Pi_{ik}\,\Pi_{jl} - \Pi_{il}\,\Pi_{jk}}{h}
  + R_{ijkl}\,,
\quad
\rho_{ijk}
= -D_i \left(\frac{\Pi_{jk}}{\sqrt h}\right)
  + D_j \left(\frac{\Pi_{ik}}{\sqrt h}\right)\,.
\label{eq:rho}
\end{equation}

\subsection{\label{sec:eom}
Hamilton's equations and the algebra of constraints
}

The equations of motion consist first in two sets of constraint
 equations, $ \delta\mathcal H^*/\delta N = 0 $ and
 $ \delta\mathcal H^*/\delta\beta^i = 0 $\,, which are
\begin{equation}
\mathcal C = 0\,,
\quad
\mathcal C_i = 0\,.
\label{eq:scons1}
\end{equation}
The equation of motion for $ \Omega_{ij} $\,,
 $ \delta\mathcal H^*/\delta\Omega_{ij} = 0 $\,, is also a constraint:
\begin{equation}
2\,\Psi^{ij}
+ \frac{\partial f}{\partial\Omega_{ij}}
= 0\,.
\label{eq:scons2}
\end{equation}
Depending on the function $ f(\varrho_{\mu\nu\rho\sigma}) $\,, this equation
 may or may not be invertible to give all the components of
 $ \Omega_{ij} $ in terms of $ \Psi^{ij} $ as well as $ h_{ij} $\,,
 $ \Pi_{ij} $ and their spatial derivatives.
As announced in the Introduction, the number of extra degrees of
 freedom beyond those of General Relativity will depend on the
 invertibility properties of \eqref{eq:scons2}.\footnote{
Notice that nonlinearity of $ f $ in $ \Omega_{ij} =
 \mathcal R_{i\mathbf nj\mathbf n} $ is essential for the
 invertibility of \eqref{eq:scons2} as well as for the appearance of
 fourth-order time-derivatives in \eqref{eq:eom0}.
}
We shall come back to this issue at the end of this Section.
A systematic way to reduce the action using \eqref{eq:scons2} will be
 presented in Section~\ref{sec:red} on some specific examples.

As for the dynamical equations, the first set, $ \dot h_{ij} =
 \delta\mathcal H^*/\delta p^{ij} $ and $ \dot\Psi^{ij} =
 \delta\mathcal H^*/\delta\Pi_{ij} $\,, gives the velocities in terms of
 the canonical variables:
\begin{equation}
\left\{
\begin{aligned}
\dot h_{ij}
& = \pounds_\beta h_{ij} - 2\,N\,\frac{\Pi_{ij}}{\sqrt h}\,, \\
\dot\Psi^{ij}
& = \pounds_\beta \Psi^{ij}
    + \frac{N}{\sqrt h}\,
      (\Pi\,\Psi^{ij}
       + (\Psi \cdot \Pi)\,h^{ij}
       + 2\,(\Psi \cdot \Pi)^{(ij)}
       - 2\,p^{ij})
    - 2\,N\,\frac{\Pi_{kl}}{\sqrt h}\,\frac{\partial f}{\partial\rho_{ikjl}}
    - D_k \left(N\,\frac{\partial f}{\partial\rho_{k(ij)}}\right)\,.
\end{aligned}
\right.
\label{eq:ceom1}
\end{equation}
The second set, $ \dot p^{ij} = -\delta\mathcal H^*/\delta h_{ij} $ and
 $ \dot\Pi_{ij} = -\delta\mathcal H^*/\delta\Psi^{ij} $ yields, using
 \eqref{eq:scons1}:
\begin{equation}
\left\{
\begin{aligned}
\dot p^{ij}
& = \sqrt h\,\pounds_\beta \left(\frac{p^{ij}}{\sqrt h}\right)
    + \frac{N}{\sqrt h}\,
      [(\Psi \cdot \Pi)\,\Pi^{ij}
       + \Psi_{kl}\,\Pi^{ik}\,\Pi^{jl}
       + h^{ij}\,
         ((\Psi \cdot \Pi)\,\Pi
          + \Psi \cdot \Pi \cdot \Pi
          - 2\,p \cdot \Pi)] \\
& \quad
    + \frac{\sqrt h}{2}\,
      [D_k (\Psi^{ij}\,\partial^k N - 2\,\Psi^{k(i}\,\partial^{j)} N)
       + h^{ij}\,(N\,D_{kl} \Psi^{kl} - \Psi^{kl}\,D_{kl} N)]
    + \frac{\sqrt h}{2}\,
      \frac{\delta(N\,f)}{\delta h_{ij}}\,, \\
\dot\Pi_{ij}
& = \sqrt h\,\pounds_\beta \left(\frac{\Pi_{ij}}{\sqrt h}\right)
    - \frac{N}{\sqrt h}\,(\Pi\,\Pi_{ij} + (\Pi \cdot \Pi)_{ij})
    + \sqrt h\,(N\,\Omega_{ij} - D_{ij} N)\,,
\end{aligned}
\right.
\label{eq:ceom2}
\end{equation}
where the last term of the right-hand side of the first equation
 reads, using \eqref{eq:scons2},
\begin{equation}
\begin{aligned}
\frac{\delta(N\,f)}{\delta h_{ij}}
& = -N\,R^{(i}{}_{klm}\,\frac{\partial f}{\partial\rho_{j)klm}}
    + 2\,D_{kl} \left(N\,\frac{\partial f}{\partial\rho_{k(ij)l}}\right)
    - 4\,N\,\frac{\Pi^{(i}{}_l\,\Pi_{km}}{h}\,
      \frac{\partial f}{\partial\rho_{j)klm}}
    - 2\,N\,h^{ij}\,\frac{\Pi_{km}\,\Pi_{ln}}{h}\,
      \frac{\partial f}{\partial\rho_{klmn}} \\
& \quad
    + N 
      \left[
       D^{(i} \left(\frac{\Pi_{kl}}{\sqrt h}\right)
       - D_k \left(\frac{\Pi^{(i}{}_l}{\sqrt h}\right)
      \right]\,
      \frac{\partial f}{\partial\rho_{j)kl}}
    - \frac{\Pi^{(i}{}_l}{\sqrt h}\,
      D_k \left(N\,\frac{\partial f}{\partial\rho_{|kl|j)}}\right)
    - h^{ij}\,\frac{\Pi_{lm}}{\sqrt h}\,
      D_k \left(N\,\frac{\partial f}{\partial\rho_{klm}}\right) \\
& \quad
    - D_l
      \left(
       N\,\frac{\Pi^{(i}{}_k}{\sqrt h}\,\frac{\partial f}{\partial\rho_{j)kl}}
       + N\,\frac{\Pi^l{}_k}{\sqrt h}\,\frac{\partial f}{\partial\rho_{k(ij)}}
      \right)
    + 2\,N\,(\Psi \cdot \Omega)^{(ij)}
\end{aligned}
\end{equation}
up to surface integrals which we omit.

We have checked that, as they must, the two constraint equations
 \eqref{eq:scons1} with \eqref{eq:scons2} reproduce the $ (00) $ and
 the $ (0i) $ components of the vacuum Euler--Lagrange equation of
 motion \eqref{eq:eom0} and that \eqref{eq:ceom1} and \eqref{eq:ceom2}
 are nothing but their $ (ij) $ components modulo constraints (the
 calculation is fairly involved and the details will be presented
 elsewhere \cite{Sendouda:prep}).

To compute the Poisson brackets of the secondary constraints
 \eqref{eq:C}, it is useful to define smeared quantities by
\begin{equation}
H[\nu]
\equiv
  \int_{\Sigma_t}\!\mathrm d^{D-1}x\,\mathcal C\,\nu\,,
\quad
M[\xi^i]
\equiv
  \int_{\Sigma_t}\!\mathrm d^{D-1}x\,\mathcal C_i\,\xi^i\,,
\end{equation}
where $ \nu(x^k) $ and $ \xi^i(x^k) $ are test functions.
This allows us to integrate by parts and to ignore boundary terms, as
 everywhere else in this paper.
A calculation shows that the Poisson brackets of the smeared
 Hamiltonian and momentum constraints, modulo \eqref{eq:scons2}, read
 the same as in General Relativity:
\begin{equation}
\left\{
\begin{aligned}
\{H[\nu_1],H[\nu_1]\}
& = M[\nu_1\,\partial^i \nu_2 - \nu_2\,\partial^i \nu_1]\,, \\
\{M[\xi_1^i],M[\xi_2^i]\}
& = M[\xi_1^j D_j \xi_2^i - \xi_2^j D_j \xi_1^i]\,, \\
\{M[\xi^i],H[\nu]\}
& = H[\pounds_\xi \nu]\,.
\end{aligned}
\right.
\label{eq:HMalgebra}
\end{equation}
Here it may be worth making a few comments on the relation between
 the constraint \eqref{eq:scons2} and the Hamiltonian and momentum
 constraints.

If \eqref{eq:scons2} is invertible with respect to $ \Omega_{ij} $\,,
 we may eliminate $ \Omega_{ij} $ from the action completely, leaving
 no further constraints.
Then the remaining constraints are the Hamiltonian and momentum
 constraints, which are of ``first-class''
 \cite{Dirac:2001,Henneaux:1992ig} representing the diffeomorphism
 invariance of the $ f(\text{Riemann}) $ action.

On the other hand, if \eqref{eq:scons2} is not completely invertible,
 some of the components give rise to non-trivial extra ``primary''
 constraints on the dynamical variable $ \Psi^{ij} $\,.
Then time derivatives of these extra constraints may give rise to
 ``secondary'' constraints \cite{Dirac:2001}.
After all the extra constraints, irrespective of primary or secondary,
 are spelled out, one can classify them into first-class and
 second-class.
In most cases, these constraints will be of second-class, which may be
 inserted in the action to reduce the dynamical degrees of freedom
 \cite{Faddeev:1988qp}.
Then, for consistency, the Hamiltonian and momentum constraints
 expressed in terms of the reduced phase space variables should also
 satisfy Eq.~\eqref{eq:HMalgebra}.
As examples, this will be confirmed below in the case of Einstein
 gravity as well as of $ f(\mathcal R) $ gravity.

In some exceptional cases, these extra constraints may happen to be of
 first-class, that is, the action may acquire a larger gauge
 invariance:
It is known that the ``$ \text{Weyl}^2 $'' action in $ D = 4 $ has a
 conformal invariance and in that case the constraint algebra of the
 Poisson brackets is extended to incorporate the generator of conformal
 transformations \cite{Boulware:1983yj}.

Now that we have completed the presentation of the general formalism,
 let us turn to some ``practical'' applications.

\section{\label{sec:red}
Phase space reduction
}

In this Section, we show how the constraint \eqref{eq:scons2} is
 utilised to find ``reduced'' Hamiltonians for various sub-classes of
 $ f(\text{Riemann}) $ gravity.
We follow the procedure advocated in \cite{Faddeev:1988qp}:
Second-class constraints arising from \eqref{eq:scons2} are inserted
 into the first-order action \eqref{eq:L*2} to eliminate as many
 components of $ \Omega_{ij} $ and of dynamical variables as possible.
If there remain any constraints on $ \Psi^{ij} $\,, irrespective of
 whether they give rise to further constraints on the other dynamical
 variables or not, it becomes necessary to redefine canonical momenta
 conjugate to the reduced sets of $ h_{ij} $ and $ \Psi^{ij} $\,.
As we shall see below, they are relatively easily read off from the
 first-order action.
Consequently the Hamiltonian in terms of the reduced set of variables
 is obtained.

\subsection{
$ f = \mathcal R_{\mu\nu\rho\sigma}\,\mathcal R^{\mu\nu\rho\sigma} $
}

This is a simple ``generic'' example where all the components of
 $ \Omega_{ij} $ can be extracted from \eqref{eq:scons2}, so that there
 are no extra constraint on the dynamical variables.

The first step is to decompose $ f(\varrho_{\mu\nu\rho\sigma}) $ as
\begin{equation}
f(\varrho_{\mu\nu\rho\sigma})
= \varrho_{\mu\nu\rho\sigma}\,\varrho^{\mu\nu\rho\sigma}
= 4\,\Omega \cdot \Omega
  + \rho_{ijkl}\,\rho^{ijkl}
  - 4\,\rho_{ijk}\,\rho^{ijk}\,,
\label{eq:Riemann2f}
\end{equation}
where $ \rho_{ijkl} $ and $ \rho_{ijk} $ are given in \eqref{eq:rho}.
Thus the constraint \eqref{eq:scons2} reads
\begin{equation}
\Omega^{ij}
= -\frac{\Psi^{ij}}{4}\,.
\label{eq:Riemann2scons}
\end{equation}
This constraint is inserted into the first-order action \eqref{eq:L*2}
 to eliminate $ \Omega_{ij} $\,.
The Hamiltonian density is therefore given by \eqref{eq:H} and
 \eqref{eq:C}, where
\begin{equation}
\Psi \cdot \Omega
+ \frac{1}{2}\,f(\varrho_{\mu\nu\rho\sigma})
= -\frac{\Psi \cdot \Psi}{8}
  + \frac{1}{2}\,(\rho_{ijkl}\,\rho^{ijkl} - 4\,\rho_{ijk}\,\rho^{ijk})\,.
\end{equation}
It depends on the usual set of variables of General Relativity,
 $ \{h_{ij},p^{ij},N,\beta^i\} $\,, plus $ D\,(D-1)/2 $ extra degrees of
 freedom, $ \{\Psi^{ij},\Pi_{ij}\} $\,.

Let us mention that the form of the canonical equations of motion
 \eqref{eq:ceom1} and \eqref{eq:ceom2} is unchanged even if
 \eqref{eq:scons2} is inserted into the Hamiltonian before taking the
 variations.
This is because the Hamiltonian depends on $ \Omega_{ij} $ only through
 the combination
\begin{equation}
\Psi \cdot \Omega + \frac{1}{2}\,f(\varrho_{\mu\nu\rho\sigma})\,,
\label{eq:Omega}
\end{equation}
so that the additional terms appearing in the equations of motion due
 to the constraint \eqref{eq:scons2} are all proportional to the
 derivative of \eqref{eq:Omega} with respect to $ \Omega_{ij} $\,,
 which vanish by virtue of \eqref{eq:scons2} itself.

It is also worth mentioning here a decomposition of $ \Psi^{ij} $ in
 the ``generic'' case.
The $ D\,(D-1)/2 $ components of $ \Psi^{ij} $ can be decomposed into
 two irreducible parts:
\begin{equation}
\Phi
\equiv
  \frac{\Psi}{D-1}\,,
\quad
\psi^{ij}
\equiv
  {}_\mathbb T\Psi^{ij}\,,
\end{equation}
where $ \Psi \equiv h_{ij}\,\Psi^{ij} $ and the symbol $ \mathbb T $
 denotes the traceless part.
Noting that there is an arbitrariness in choosing canonical momenta
 conjugate to the variables $ \{h_{ij},\Phi,\psi^{ij}\} $\,, one finds it
 most convenient to introduce
\begin{equation}
\tilde p^{ij}
\equiv
  p^{ij}
  - \frac{1}{D-1}\,
    (\Pi\,\Psi^{ij} + \Psi\,{}_\mathbb T\Pi^{ij})\,,
\quad
\pi_{ij}
\equiv
  {}_\mathbb T\Pi_{ij}
\end{equation}
resulting in
\begin{equation}
\mathcal L^*
= \tilde p \cdot \dot h
  + \Pi\,\dot\Phi
  + \pi \cdot \dot\psi
  - \mathcal H^*\,,
\end{equation}
where $ \mathcal H^* $ is to be given in terms of the new variables.
Now it is manifest that the ``generic'' theory contains a scalar
 degree of freedom $ \{\Phi,\Pi\} $ and traceless tensor degrees of
 freedom $ \{\psi^{ij},\pi_{ij}\} $ having $ (D+1)\,(D-2)/2 $ components
 on top of the canonical metric degrees of freedom
 $ \{h_{ij},\tilde p^{ij}\} $\,.

\subsection{
$ f = \mathcal R $
}

We show here how the ADM Hamiltonian for General Relativity follows
 from our general formalism.

The decomposition of $ f(\varrho_{\mu\nu\rho\sigma}) $ is 
\begin{eqnarray}
f
= \varrho
\equiv
  g^{\mu\rho}\,g^{\nu\sigma}\,\varrho_{\mu\nu\rho\sigma}
= -2\,\Omega + \rho\,,
\end{eqnarray}
where $ \Omega \equiv h^{ij}\,\Omega_{ij} $ and $ \rho \equiv
 h^{ik}\,h^{jl}\,\rho_{ijkl} $ so that the primary
 constraint \eqref{eq:scons2} reads
\begin{equation}
\Psi^{ij}
= h^{ij}\,.
\label{eq:Rscons2}
\end{equation}
Hence it ``freezes'' all the extra degrees of freedom of
 $ \Psi^{ij} $\,.
Though \eqref{eq:Rscons2} does not give $ \Omega_{ij} $\,, the terms
 containing $ \Omega_{ij} $ in the Hamiltonian will all drop out due to
 the linear dependence of $ f $ on $ \Omega_{ij} $\,.

The Hamilton equations \eqref{eq:ceom1} for the velocities give a
 secondary constraint.
It tells us that, since $ \dot\Psi^{ij} = \dot h^{ij} $\,, the momentum
 conjugate to $ \Psi^{ij} $ is also frozen:
\begin{equation}
\Pi_{ij}
= 2\,p_{ij} - \frac{2\,p}{D}\,h_{ij}\,,
\end{equation}
where $ p \equiv h_{ij}\,p^{ij} $\,.

We now follow the procedure advocated in \cite{Faddeev:1988qp} and
 insert \eqref{eq:Rscons2} into the first-order action
 \eqref{eq:L*2}\,, which becomes (ignoring the divergence)
\begin{equation}
\mathcal L^*
= \tilde p \cdot \dot h - \mathcal H^*\,,
\end{equation}
with
\begin{equation}
\tilde p^{ij}\equiv
  -p^{ij} + \frac{2\,p}{D}\,h^{ij}\,.
\end{equation}
This $ \tilde p^{ij} $ plays the role of the new momentum conjugate to
 $ h_{ij} $\,.

We now gather the results, to wit, all the dynamical variables are
 expressed in terms of $ \{h_{ij},\tilde p^{ij}\} $ as
\begin{equation}
\Psi^{ij}
= h^{ij}\,,
\quad
\Pi_{ij}
= -2\,\tilde p_{ij}
  + \frac{2\,\tilde p}{D-2}\,h_{ij}\,,
\quad
p^{ij}
= -\tilde p^{ij}
  + \frac{2\,\tilde p}{D-2}\,h^{ij}\,,
\end{equation}
where $ \tilde p \equiv h_{ij}\,\tilde p^{ij} $\,, and we plug them into
 the Hamiltonian \eqref{eq:H} and \eqref{eq:C} to obtain
\begin{equation}
\mathcal H^*
= N\,\mathcal C + \beta^i\,\mathcal C_i\,,
\end{equation}
with
\begin{equation}
\mathcal C
= \frac{2}{\sqrt h}\,
  \left(\tilde p \cdot \tilde p - \frac{\tilde p^2}{D-2}\right)
  - \frac{\sqrt h}{2}\,R\,,
\quad
\mathcal C_i
= -2\,\sqrt h\,D_j \left(\frac{\tilde p^j{}_i}{\sqrt h}\right)\,.
\label{eq:RC}
\end{equation}
This is nothing but the ADM Hamiltonian for General Relativity in
 $ D $ dimensions.
As for the equations of motion \eqref{eq:ceom1} and \eqref{eq:ceom2},
 they reduce to the ADM equations, see e.g.\ \cite{Wald:1984rg}.
Moreover, the constraints \eqref{eq:RC} are first-class as is well
 known.

\subsection{
$ f = f(\mathcal R) $
}

We show here how our general formalism yields the ``Jordan-frame''
 Hamiltonian of $ f(\mathcal R) $ gravity.

As above we have $ \varrho = -2\,\Omega + \rho $ with $ \Omega \equiv
 h^{ij}\,\Omega_{ij} $ and $ \rho \equiv h^{ik}\,h^{jl}\,\rho_{ijkl} $\,,
 that is, using \eqref{eq:rho}:
\begin{equation}
\varrho
= -2\,\Omega + \frac{\Pi^2 - \Pi \cdot \Pi}{h} + R\,,
\label{eq:f(R)rho}
\end{equation}
so that Eq.~\eqref{eq:scons2} reads
\begin{equation}
\Psi^{ij}
= f'(\varrho)\,h^{ij}
\quad\Longleftrightarrow\quad
\Phi
\equiv
  \frac{h \cdot \Psi}{D-1}
= f'(\varrho)\,,
\quad
{}_\mathbb T\Psi^{ij}
= 0\,,
\label{eq:f(R)pcons}
\end{equation}
where the symbol $ \mathbb T $ denotes the traceless part.
This primary constraint tells us, first, that the trace $ \Omega $ is
 known in terms of $ \Phi $ and other variables, and that the
 traceless part of $ \Psi^{ij} $ is constrained to be zero so that
 $ \Psi^{ij} $ reduces to one scalar degree of freedom, $ \Phi $\,.
The traceless part of $ \Omega_{ij} $ automatically disappears from the
 Hamiltonian since $ \Psi \cdot \Omega = \Phi\,\Omega $ and
 $ f(\varrho) $ depends only on the trace $ \Omega $ from the
 beginning.

The traceless part of $ \Pi_{ij} $ is also constrained through the
 Hamilton equations for the velocities \eqref{eq:ceom1}.
Indeed, the traceless part of the velocities now satisfies
 $ {}_\mathbb T(\dot\Psi^{ij}) = {}_\mathbb T(\dot h^{ij})\,\Phi $ giving a
 secondary constraint
\begin{equation}
{}_\mathbb T\Pi_{ij}
= \frac{2}{\Phi}\,{}_\mathbb Tp_{ij}\,.
\label{eq:f(R)scons}
\end{equation}

In order now to find a new momentum $ \tilde p^{ij} $ conjugate to
 $ h_{ij} $\,, we again follow \cite{Faddeev:1988qp} and insert
 \eqref{eq:f(R)pcons} and \eqref{eq:f(R)scons} into the first-order
 action \eqref{eq:L*2}, which now reads (ignoring the divergence)
\begin{equation}
\mathcal L^*
= \tilde p \cdot \dot h
  + \Pi\,\dot\Phi
  - \mathcal H^*\,,
\end{equation}
with
\begin{equation}
\tilde p^{ij}
\equiv
  -p^{ij}
  + \frac{h^{ij}}{D-1}\,(2\,p - \Phi\,\Pi)\,,
\end{equation}
where $ p \equiv h_{ij}\,p^{ij} $\,.
Gathering the results:
\begin{equation}
\Psi^{ij}
= \Phi\,h^{ij}\,,
\quad
\Pi_{ij}
= \frac{1}{\Phi}\,
  \left[
   -2\,\tilde p_{ij}
   + \frac{h_{ij}}{D-1}\,(2\,\tilde p + \Phi\,\Pi)
  \right]\,,
\quad 
p^{ij}
= -\tilde p^{ij}
  + \frac{h^{ij}}{D-1}\,(2\,\tilde p + \Phi\,\Pi)\,,
\end{equation}
where $ \tilde p \equiv h_{ij}\,\tilde p^{ij} $\,, and using
 \eqref{eq:f(R)rho} and \eqref{eq:f(R)pcons}, we have 
\begin{equation}
\Psi \cdot \Omega
= \frac{1}{2}\,\Phi\,(R - \varrho)
  + \frac{D-2}{2\,(D-1)}\,\Phi\,\frac{\Pi^2}{h}
  - \frac{2}{\Phi}\,\frac{{}_\mathbb T\tilde p \cdot {}_\mathbb T\tilde p}{h}\,,
\end{equation}
where it is understood that $ \varrho $ is known in terms of $ \Phi $
 via $ f'(\varrho) = \Phi $\,.
Therefore the Hamiltonian \eqref{eq:H} becomes $ \mathcal H^* =
 N\,\mathcal C + \beta^i\,\mathcal C_i $ with
\begin{equation}
\left\{
\begin{aligned}
\mathcal C
& = \frac{2}{\sqrt h}\,
    \left[
     \frac{{}_\mathbb T\tilde p \cdot {}_\mathbb T\tilde p}{\Phi}
     + \frac{D-2}{4\,(D-1)}\,\Phi\,\Pi^2
     - \frac{\tilde p\,\Pi}{D-1}
    \right]
    + \frac{\sqrt h}{2}\,
      (\Phi\,\varrho - f(\varrho) - \Phi\,R + 2\,D_i D^i \Phi)\,, \\
\mathcal C_i
& = -2\,\sqrt h\,D_j \left(\frac{\tilde p_i{}^j}{\sqrt h}\right)
    + \Pi\,\partial_i \Phi\,.
\end{aligned}
\right.
\label{eq:f(R)C}
\end{equation}
The constraints \eqref{eq:f(R)C} give the standard Jordan-frame
 Hamiltonian for $ f(\mathcal R) $ gravity, see \cite{Deruelle:2009pu}.
Moreover, it can be checked that they are first-class.

One can also make a canonical transformation of the Hamiltonian into
 that of Einstein gravity with a minimally coupled scalar field.
These metrics are related by a $ D $-dimensional conformal
 transformation, see e.g.\ \cite{Maeda:1988ab} and
 \cite{Deruelle:2009pu}.

\subsection{
$ f = \mathcal C_{\mu\nu\rho\sigma}\,\mathcal C^{\mu\nu\rho\sigma} $
}

We consider here the case when $ f =
 \mathcal C_{\mu\nu\rho\sigma}\,\mathcal C^{\mu\nu\rho\sigma} $ where
 $ \mathcal C_{\mu\nu\rho\sigma} $ is the $ D $-dimensional Weyl tensor.
In terms of the Riemann and Ricci tensors, it is expressed as
\begin{equation}
f(\mathcal R_{\mu\nu\rho\sigma})
= \mathcal R_{\mu\nu\rho\sigma}\,\mathcal R^{\mu\nu\rho\sigma}
 - \frac{4}{D-2}\,\mathcal R_{\mu\nu}\,\mathcal R^{\mu\nu}
 + \frac{2}{(D-1)\,(D-2)}\,\mathcal R^2\,.
\end{equation}
This theory yields conformally invariant equations of motion in
 $ D = 4 $\,.

To decompose $ f(\varrho_{\mu\nu\rho\sigma}) $\,, we need
\begin{equation}
\left\{
\begin{aligned}
\varrho_{\mu\nu\rho\sigma}\,\varrho^{\mu\nu\rho\sigma}
& = 4\,\Omega \cdot \Omega
    + \rho_{ijkl}\,\rho^{ijkl}
    - 4\,\rho_{ijk}\,\rho^{ijk}\,, \\
\varrho_{\mu\nu}\,\varrho^{\mu\nu}
& = \Omega^2
    + \Omega \cdot \Omega
    - 2\,\rho \cdot \Omega
    + \rho \cdot \rho
    - 2\,\rho_i\,\rho^i\,, \\
\varrho^2
& = (-2\,\Omega + \rho)^2\,,
\end{aligned}
\right.
\end{equation}
where $ \Omega \equiv h^{ij}\,\Omega_{ij} $\,, $ \rho_{ij} \equiv
 h^{kl}\,\rho_{ikjl} $\,, $ \rho \equiv h^{ik}\,h^{jl}\,\rho_{ijkl} $ and
 $ \rho_i \equiv h^{jk}\,\rho_{jik} $ with
\begin{equation}
\left\{
\begin{aligned}
&
\rho_{ijkl}
= \frac{\Pi_{ik}\,\Pi_{jl} - \Pi_{il}\Pi_{jk}}{h} + R_{ijkl}\,,
\quad
\rho_{ij}
= \frac{\Pi\,\Pi_{ij} - (\Pi \cdot \Pi)_{ij}}{h} + R_{ij}\,,
\quad
\rho
= \frac{\Pi^2 - \Pi \cdot \Pi}{h} + R\,, \\
&
\rho_{ijk}
= -D_i \left(\frac{\Pi_{jk}}{\sqrt h}\right)
  + D_j \left(\frac{\Pi_{ik}}{\sqrt h}\right)\,,
\quad
\rho_i
= -D^j \left(\frac{\Pi_{ij}}{\sqrt h}\right)
  + \partial_i \left(\frac{\Pi}{\sqrt h}\right)\,.
\end{aligned}
\right.
\end{equation}
For comparison with the Riemann squared case, see \eqref{eq:Riemann2f}
 and \eqref{eq:rho}.
The constraint equation \eqref{eq:scons2} now reads
\begin{equation}
\Psi^{ij}
= -\frac{4}{D-2}\,
  [(D-3)\,{}_\mathbb T\Omega^{ij} + {}_\mathbb T\rho^{ij}]\,,
\end{equation}
where again $ \mathbb T $ denotes the traceless part.
Thus only the traceless part of $ \Omega_{ij} $ is determined.
As a consequence, there appears an extra primary constraint
 $ h \cdot \Psi = 0 $\,, that is, the trace part of $ \Psi^{ij} $
 disappears from the action.
This implies the number of degrees of freedom in $ \Psi^{ij} $ will be
 reduced by one.

Now we have a relation between the velocities $ \dot h \cdot \Psi
 + h \cdot \dot\Psi = 0 $ as a consequence of the above primary
 constraint so that the Hamilton equations for the velocities
 \eqref{eq:ceom1} yield a secondary constraint,
\begin{equation}
p - \frac{D}{2}\,{}_\mathbb T\Psi \cdot {}_\mathbb T\Pi
= 0\,,
\label{eq:W2scons}
\end{equation}
where $ p \equiv h_{ij}\,p^{ij} $\,.
In contradistinction with the previous example of $ f(\mathcal R) $
 gravity, this equation does not constrain $ \Pi $ but $ p $\,.
For the moment, we decide not to insert this constraint into the
 action.
The reason will be clarified below.

In order now to find the new momenta $ \tilde p^{ij} $ and $ \pi_{ij} $
 conjugate to $ h_{ij} $ and $ \psi^{ij} \equiv {}_\mathbb T\Psi^{ij} $\,,
 respectively, we again follow \cite{Faddeev:1988qp} and write the
 first-order action \eqref{eq:L*2} as (ignoring the divergence)
\begin{equation}
\mathcal L^*
= \tilde p \cdot \dot h + \pi \cdot \dot\psi - \mathcal H^*\,,
\label{eq:W2L}
\end{equation}
with
\begin{equation}
\tilde p^{ij}
\equiv
  p^{ij} - \frac{\Pi}{D-1}\,\psi^{ij}\,,
\quad
\pi_{ij}
\equiv
  {}_\mathbb T\Pi_{ij}\,.
\end{equation}
Let us gather the results:
\begin{equation}
\Psi^{ij}
= \psi^{ij}\,,
\quad
\Pi_{ij}
= \pi_{ij} + \frac{\Pi}{D-1}\,h_{ij}\,,
\quad 
p^{ij}
= \tilde p^{ij} + \frac{\Pi}{D-1}\,\psi^{ij}\,,
\end{equation}
where we see that $ \Pi $ is not determined.
Plugging these into the Hamiltonian \eqref{eq:H} one finds
\begin{equation}
\mathcal H^*
= N\,\mathcal C
  + \beta^i\,\mathcal C_i
  + \frac{2}{D-1}\,\frac{N\,\Pi}{\sqrt h}\,\mathcal C_W\,,
\label{eq:W2H}
\end{equation}
with
\begin{equation}
\left\{
\begin{aligned}
\mathcal C
& = \frac{1}{\sqrt h}\,
    \left(
     \frac{D-4}{D-3}\,\psi \cdot \pi \cdot \pi
     - 2\,\tilde p \cdot \pi
    \right)
    + \sqrt h\,
      \left[
       \frac{D-2}{8\,(D-3)}\,\psi \cdot \psi
       + D_{ij} \psi^{ij}
       + \frac{1}{D-3}\,\psi \cdot {}_\mathbb TR
      \right] \\
& \quad
    - \frac{\sqrt h}{2}\,
      \left(
       {}_\mathbb T\rho_{ijkl}\,{}_\mathbb T\rho^{ijkl}
       - 4\,{}_\mathbb T\rho_{ijk}\,{}_\mathbb T\rho^{ijk}
      \right)\,, \\
\mathcal C_i
& = -2\,\sqrt h\,D_j \left(\frac{\tilde p_i{}^j}{\sqrt h}\right)
    + \pi_{jk}\,D_i \psi^{jk}
    + 2\,\sqrt h\,
      D_k \left(\frac{\pi_{ij}\,\psi^{jk}}{\sqrt h}\right)\,, \\
\mathcal C_W
& = \frac{D}{2}\,\psi \cdot \pi - \tilde p\,,
\end{aligned}
\right.
\end{equation}
where $ \tilde p \equiv h_{ij}\,\tilde p^{ij} $\,.
We note that the Hamiltonian is linear in $ \Pi $\,.
In other words, $ \Pi $ acts as a Lagrange multiplier whose equation
 of motion, $ \mathcal C_W = 0 $\,, coincides with the secondary
 constraint \eqref{eq:W2scons}.

Boulware \cite{Boulware:1983yj} showed that, in $ D = 4 $\,, the
 Poisson bracket algebra of the constraints closes, that is,
 $ \mathcal C_W $ as well as $ \mathcal C $ and $ \mathcal C_i $ are
 first-class constraints.
Boulware also showed that $ \mathcal C_W $ is the generator of
 conformal transformations under which the theory is invariant.
Thus, to summarise, the Lagrangian for the Weyl squared theory in
 $ D = 4 $ keeps the original form \eqref{eq:W2L},
\begin{equation}
\mathcal L^*
= \tilde p \cdot \dot h + \pi \cdot \dot\psi - \mathcal H^*\,,
\end{equation}
where $ \mathcal H^* $ is given by \eqref{eq:W2H}, but with all the
 constraints $ \mathcal C $, $ \mathcal C_i $ and $ \mathcal C_W $
 being first-class.

In passing, we note that it may be more transparent to separate out
 the determinant of $ h_{ij} $ as an independent canonical variable.
Namely, in addition to $ \tilde p \equiv h_{ij}\,\tilde p^{ij} $\,,
 introducing
\begin{eqnarray}
\eta
\equiv
  \ln h^{1/(D-1)}\,,
\quad
\hat h_{ij}
\equiv
  \mathrm e^{-\eta}\,h_{ij}\,,
\quad
\hat p^{ij}
\equiv
  \mathrm e^\eta\,{}_\mathbb T\tilde p^{ij}\,,
\end{eqnarray}
we express the Lagrangian as
\begin{eqnarray}
\mathcal L^*
= \hat p \cdot \dot{\hat h}
  + \tilde p\,\dot\eta
  + \pi \cdot \dot\psi
  - \mathcal H^*\,,
\end{eqnarray}
where $ \mathcal H^* $ is given by
\begin{equation}
\mathcal H^*
= N\,\mathcal C + \beta^i\,\mathcal C_i + W\,\mathcal C_W\,.
\end{equation}
Here we have replaced $ \Pi $ by $ W \equiv
 2\,(D-1)^{-1}\,N\,\sqrt h^{-1}\,\Pi $\,, and $ \mathcal C $\,,
 $ \mathcal C_i $ and $ \mathcal C_W $ are to be expressed in terms of
 the new canonical variables $ \{\hat h_{ij},\hat p^{ij}\} $\,,
 $ \{\eta,\tilde p\} $ and $ \{\psi^{ij},\pi_{ij}\} $\,.

In $ D > 4 $\,, however, the Poisson brackets of $ \mathcal C_W $ with
 $ \mathcal C $ and $ \mathcal C_i $ no longer close, giving rise to
 additional secondary constraints.
This is because, although the $ D $-dimensional Weyl tensor,
 $ \mathcal C^\mu{}_{\nu\rho\sigma} $\,, is conformally invariant in any
 dimensions, the action is not conformally invariant any longer in
 dimensions other than $ D = 4 $\,.
In this case, there may be more secondary constraints from the time
 derivatives of these secondary constraints.
They will be all second-class constraints.
Inserting them into the action will reduce the phase space
 considerably.
Because of rather involved calculations, we have not checked how many
 second-class constraints would appear in the end.
We plan to come back to this issue in a future publication
 \cite{Sendouda:prep}.

\section{
Conclusion
}

We have presented in this paper a canonical formulation of
 $ f(\text{Riemann}) $ theories in a form as compact as possible.
It includes in a unifying manner well-known subcases and should prove
 useful to analyse the properties of various theories of ``extended
 gravity,'' such as the global charges associated with the solutions,
 their (in)stability and the (non-)positivity of energy or the
 junction conditions.

\begin{acknowledgments}
ND thanks the Yukawa Institute for Theoretical Physics for its
 hospitality.
MS and YS thank the hospitality of APC at Paris 7 and ASC at LMU,
 where this work was completed.
The work of MS is supported in part by JSPS Grant-in-Aid for
 Scientific Research (A) No.~18204024, (A) No.~21244033, and by JSPS
 Grant-in-Aid for Creative Scientific Research No.~19GS0219.
YS and DY are supported by MEXT through Grant-in-Aid for JSPS Fellows
 No.~19-7852 (YS) and No.~20-1117 (DY).
This work is also supported in part by the Grant-in-Aid for the Global
 COE Program ``The Next Generation of Physics, Spun from Universality
 and Emergence'' from MEXT.
\end{acknowledgments}

\appendix
\section{
The Ostrogradsky Hamiltonian of $ f(\text{Riemann}) $ gravity
}

For completeness we give here the canonical transformation which
 relates the variables used in this paper to the ``Ostrogradsky'' one,
 that is, the extrinsic curvature of the ADM foliation, see
 \cite{Demaret:1998dm} for an Ostrogradsky treatment of quadratic
 theories.

The Ostrogradsky action written in terms of $ \varsigma_{\mu\nu\rho\sigma}
 \equiv \mathcal R_{\mu\nu\rho\sigma}|_{K_{ij}\to Q_{ij}} $\,, where $ Q_{ij} $ is
 independent of other fields, is
\begin{equation}
S
= \int_\mathcal M\!\mathrm d^Dx\,\sqrt{-g}\,
  \left[
   \frac{1}{2}\,f(\varsigma_{\mu\nu\rho\sigma})
   + 2\,u^{ij}\,(K_{ij} - Q_{ij})
  \right].
\end{equation}
The momenta canonically conjugate to $ h_{ij} $ and $ Q_{ij} $\,,
 respectively, are
\begin{equation}
k^{ij}
\equiv
  \frac{\delta S}{\delta\dot h_{ij}}
= \sqrt h\,u^{ij}\,,
\quad
P^{ij}
\equiv
  \frac{\delta S}{\delta\dot Q_{ij}}
= -\frac{\sqrt h}{2}\,
  \frac{\partial f}{\partial\Sigma_{ij}}\,,
\end{equation}
where $ \Sigma_{ij} \equiv n^\mu\,n^\nu\,\varsigma_{i\mu j\nu} $\,.
We have the following relations with the variables of this paper:
\begin{equation}
\Pi_{ij}
= -\sqrt h\,Q_{ij}\,,
\quad
\Psi^{ij}
= \,\frac{P^{ij}}{\sqrt h}\,.
\end{equation}
They give
\begin{equation}
p \cdot \dot h + \Pi \cdot \dot\Psi
= k \cdot \dot h
  + P \cdot \dot Q
  - \frac{\mathrm d}{\mathrm dt} (Q \cdot P)
\end{equation}
with
\begin{equation}
k^{ij}
= p^{ij} + \frac{Q \cdot P}{2}\,h^{ij}\,.
\end{equation}
Plugging these relations into our Hamiltonian \eqref{eq:H}, we obtain
 the Ostrogradsky Hamiltonian $ \mathcal H^* = N\,\mathcal C +
 \beta^i\,\mathcal C_i $ with
\begin{equation}
\left\{
\begin{aligned}
\mathcal C
& = Q \cdot Q \cdot P
    + 2\,Q \cdot k
    - P \cdot \Sigma
    + \sqrt h\,D_{ij} \left(\frac{P^{ij}}{\sqrt h}\right)
    - \frac{\sqrt h}{2}\,f(\varsigma_{\mu\nu\rho\sigma})\,, \\
\mathcal C_i
& = -2\,\sqrt h\,D_j \left(\frac{k_i{}^j}{\sqrt h}\right)
    + P^{jk}\,D_i Q_{jk}
    - 2\,\sqrt h\,D_k \left(\frac{Q_{ij}\,P^{jk}}{\sqrt h}\right)\,,
\end{aligned}
\right.
\end{equation}
where $ \varsigma_{\mu\nu\rho\sigma} $ is expressed in terms of $ h_{ij} $\,,
 $ Q_{ij} $ and $ \Sigma_{ij} $\,, and where $ \Sigma_{ij} $ is
 constrained to satisfy
\begin{equation}
\frac{\partial f}{\partial\Sigma_{ij}}
= -2\,\frac{P^{ij}}{\sqrt h}\,.
\end{equation}

\end{document}